# Recent insights into the impact of geopolitical tensions: Quantifying the structure of computer science professors of Chinese descent in the United States

Yongzhen Wang[1,2]

yongzhenwang@dlut.edu.cn

[1]School of Public Administration and Policy, Dalian University of Technology, Dalian 116024, China

[2]WISE Lab, Dalian University of Technology, Dalian 116024, China

## Abstract

The geopolitical tensions between China and the US have dramatically reshaped the American scientific workforce's landscape. To gain a deeper understanding of this circumstance, this study selects the discipline of computer science as a representative case for empirical investigations, aiming to explore the current situation of US-based Chinese-descent computer science professors. One thousand and seventy-eight tenured or tenure-track professors of Chinese descent from the computer science departments of 108 prestigious US universities are profiled, in order to quantify their structure primarily along gender, schooling, and expertise lines. The findings presented in this paper suggest that China-US tensions have made it more difficult for the US higher education system to retain valuable computer science professors of Chinese descent, particularly those in their mid-to late career stages, and that nearly 50% of the existing professors have less than seven years of faculty experience. In addition, the deterioration in faculty retention varies across fields of research, education backgrounds, and gender groups. Specifically, among the professors we are concerned about, those who do not work on AI or Systems, those who lack study experience at US universities, and those who are women, are underrepresented, albeit in different forms and to varying degrees. In a nutshell, the focal professoriate has not only shrunk in size, as has been widely reported, but also lost some of its diversity in structure. This paper has policy implications for the mobility of scientific talent, especially in an era of geopolitical challenges.

## Keywords

# Introduction

Science is never isolated from politics, and it is often affected by national and international policies. On the negative side, the geopolitical tensions between China and the United States (US) in recent years have strikingly exemplified some of the adverse consequences that may arise from the politicization of science (Flynn, Glennon, Murciano-Goroff, & Xiao, 2024; Mok, Shen, & Gu, 2024; Tang, 2024; Wang, 2024). As a case in point, the China Initiative——an effort which was implemented by the US Department of Justice and aimed at prosecuting perceived Chinese spies in American research and industry in order to combat so-called economic espionage (Choi, 2021, June 14)——has been heavily criticized for its racial overtones and has led to a series of unfavorable outcomes that have clouded both countries' prospects in the realm of science (Aghion et al., 2023; Jia, Roberts, Wang, & Yang, 2024; Li & Wang, 2024; Xie, Lin, Li, He, & Huang, 2023). Although the China Initiative has officially terminated in 2022, extensive investigations into foreign influence by US federal funding agencies (e.g., National Institutes of Health) are still ongoing; meanwhile, US universities' policies with respect to collaborations with Chinese scientists are in constant flux (Jia et al., 2024; Samson, 2024, September 3).

China-US tensions have dramatically reshaped the American scientific workforce's landscape. Evidence suggests that the annual number of Chinese-descent scientists leaving the US has increased significantly since the China Initiative was launched in 2018, in large part due to widespread feelings of fear and anxiety among this group (Xie et al., 2023). Their departure has, to a certain extent, reversed the long-standing trend of international academic mobility——for decades the US has been regarded as the preferred destination for international students and scientists seeking advanced education and research opportunities, especially in STEM areas (Ganguli, Kahn, & MacGarvie, 2020; Rovito, Kaushik, & Aggarwal, 2021; Walsh, 2015). Undoubtedly, the influence of the reverse brain drain on the scientific community is profound and multifaceted, not least because it has abruptly altered the structure of the American professoriate who are at the forefront of preparing the next generation of scientists. As a matter of fact, the US remains the world's powerhouse in higher education, with data showing that approximately 13% of full-time faculty at US degree-granting postsecondary institutions identify as Asian (National Center for Education Statistics, 2024). Nevertheless, the magnitude and implications of the structural change in the American professoriate are far from clear.

This study is among the few that attempt to explore the current situation of US-based Chinese-descent professors, with the purpose of providing additional insights into the impact of China-US tensions on science as well as on science education. Here, the discipline of computer science is selected as a representative case for empirical investigations. The reasons are three-fold as follows. First, the US has historically been a leader in computer science, with its universities dominating computer science education, consistently at the top globally. Second, China has emerged as the most important foreign supplier of computer scientists to the US since the early 21st century (Finocchi, Ribichini, & Schaerf, 2023). Third, amid the escalating geopolitical tensions, US-based Chinese-descent researchers working in computer science have exhibited the highest percentage of feeling insecure and unwelcome, i.e., they faced higher incentives to leave the US (Xie et al., 2023). Briefly speaking, this study necessitates the creation of a sample that comprises the profiles of computer science professors of Chinese descent in the US. For the sake of feasibility, only those who are employed by top-tier universities and hold tenured or tenure-track positions (so they can individually advise students) are considered in this study. This cross-sectional sample provides us with a snapshot of that population, enabling a quantitative analysis of their personal characteristics in the context of China-US tensions. The only work closely relevant to this study is presented at Drafty (accessed via [https://drafty.cs.brown.edu/csprofessors](https://drafty.cs.brown.edu/csprofessors)), which is a publicly-editable data repository pertaining to thousands of computer science professors working at American and Canadian universities. However, due to its crowdsourcing nature and flexible entry criteria for data, the sample created for this study is not originated from this repository.

Drawing upon the created cross-sectional sample, this study seeks to quantify the structure of computer science professors of Chinese descent in the US, primarily from three perspectives: research field, education background, and gender stratification, by which many intriguing yet open questions can be addressed. For example, is there evidence that some professors suffered more from China-US tensions? If yes, who are they? And can we give an unambiguous interpretation for this disproportionate impact? Based on the empirical investigations, an extended discussion is presented at the end of this paper to not only inform efforts aimed at improving the conditions of US-based Chinese-descent scientific workers, but also highlight the importance of an open research environment in an era of geopolitical challenges. Additionally, this study adds to the existing knowledge on international academic mobility, emphasizing mainland China's universities as a significant contributor to the discipline of computer science.

## Data collection and processing

This study examines tenured or tenure-track professors of Chinese descent from the computer science departments of 108 well-known US universities (refer to Table A1 for more details, in **Appendix**), excluding non-core faculty such as lecturers, instructors, professors of practice, clinical, adjunct, affiliate, teaching, and research professors. All of the selected universities are among the top 100 in the US for computer science, with reference to two popular ranking systems by USNews.com and Dr. Emery Berger (called CS Rankings, accessed via https://csrankings.org/) as well as a recently proposed one (called CS Open Rankings, accessed via https://drafty.cs.brown.edu/csopenrankings/) by Dr. Jeff Huang et al. who take university reputation, faculty productivity, student placement, and academic recognition into account. For some universities without pure computer science departments, multidisciplinary or interdisciplinary departments (e.g., "department of electrical engineering and computer science") or higher-level divisions related to computer science (e.g., "school of computing") are targeted instead, if any; however, faculty members who are not directly engaged in computer science within these alternative institutions are not included. Among those eligible, individuals of Chinese descent were manually identified through their self- or publicly disclosed information or their Chinese surnames, yielding a sample of size $n=1078$. Certainly, newly appointed professors for 2024 are expected to be underrepresented because of their recency. Each included professor was then annotated by their gender, faculty rank, research field, bachelor's and doctoral alma maters, year of receiving their doctorate, and year of starting their faculty career in US academia. The annotation process was conducted from July 1, 2024 to August 15, 2024, using information collected from various sources, including curriculum vitae, personal webpages, publicly accessible LinkedIn profiles, ORCID profiles, and Google Scholar profiles, etc.

All annotation tasks were completed except for a few professors ($n=58$) whose bachelor's alma maters remained unknown. Specifically, self-reported genders were used when available, and otherwise humanly annotated (as men or female) on the basis of third-party reports from corresponding university websites. As for the levels of professors, the standard three-tier classification of faculty ranks——assistant, associate, and full professor——was used for ease of annotation. Moreover, each full professor was checked to see if he/she had been elected as a fellow of the ACM (Association for Computing Machinery) or a fellow of the IEEE (Institute of Electrical and Electronics Engineers), both of which are professional accolades for computer scientists worldwide, in order to further unfold the hierarchy of US-based Chinese-descent computer science professors. To facilitate comparisons of the focal professoriate across fields of research, each professor was grouped into one of four broad categories: AI (Artificial Intelligence), Interdisciplinary, Systems, and Theory, according to the CSRankings.org's taxonomy (refer to Table A2 for more details, in **Appendix**). Note that for a professor working in multiple fields, the primary one (chosen out of the four) in which he/she published most of his/her work so far was labeled as his/her specialty. Finally, if a professor held more than one undergraduate or doctoral degree, only the degree related to computer science was considered. The correlation matrix of annotated data can be found in Table A3 (in **Appendix**).

## Results and analysis

**Nearly 50% of the existing professors having ≤7 years of faculty experience**

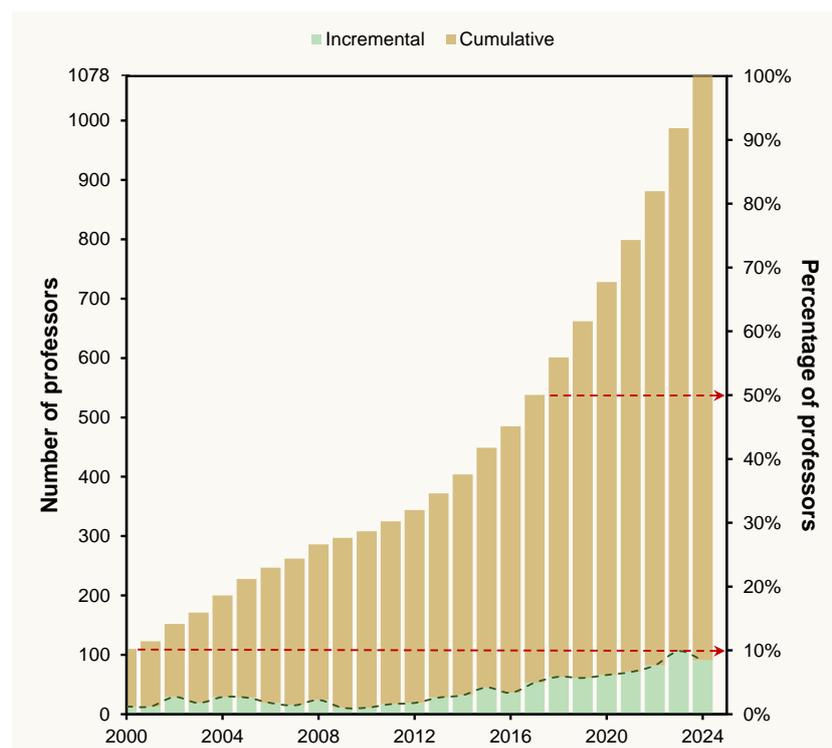

**Fig. 1** Employment dynamics: incremental (green) vs. cumulative (yellow)

Fig. 1 presents the employment dynamics of 1078 US-based Chinese-descent computer science professors, from both incremental and cumulative perspectives. Overall, from the year 2000 onwards, a continuous increase can be seen in the cumulative number, with an average annual growth rate of around 10%. Comparatively, the trend of the increments is more complex, reaching a valley point in 2010 and then starting to increase steadily, albeit with some fluctuations. This observation echoes the findings by Xie et al. (2023) that the annual number of Chinese-descent scientists migrating out of the US has gained momentum since 2010. Additionally, it can be seen that about 90% of the professors entered US academia after 2000, and that about 50% after 2018 when the China Initiative came into effect. In other words, almost half of the existing professors were appointed within the past seven years, indicating that the career age of the whole population is, on average, considerably young (<10 years). However, in fact, since at least as early as twenty years ago, China has been the most important supplier of computer scientists to the US (Finocchi et al., 2023), largely attributed to the growing number of Chinese students pursuing advanced degrees in the US and choosing to work there after graduation (Gaulé & Piacentini, 2015; Rovito et al., 2021). Together, even taking faculty turnover into account, it is reasonable and convincing to confirm the damage caused by China-US tensions to the focal professoriate, particularly those in their mid-to late career stages.

**AI retaining professors at a higher rate in recent years**
**Systems allowing for a shorter preparation time to attain a faculty position**

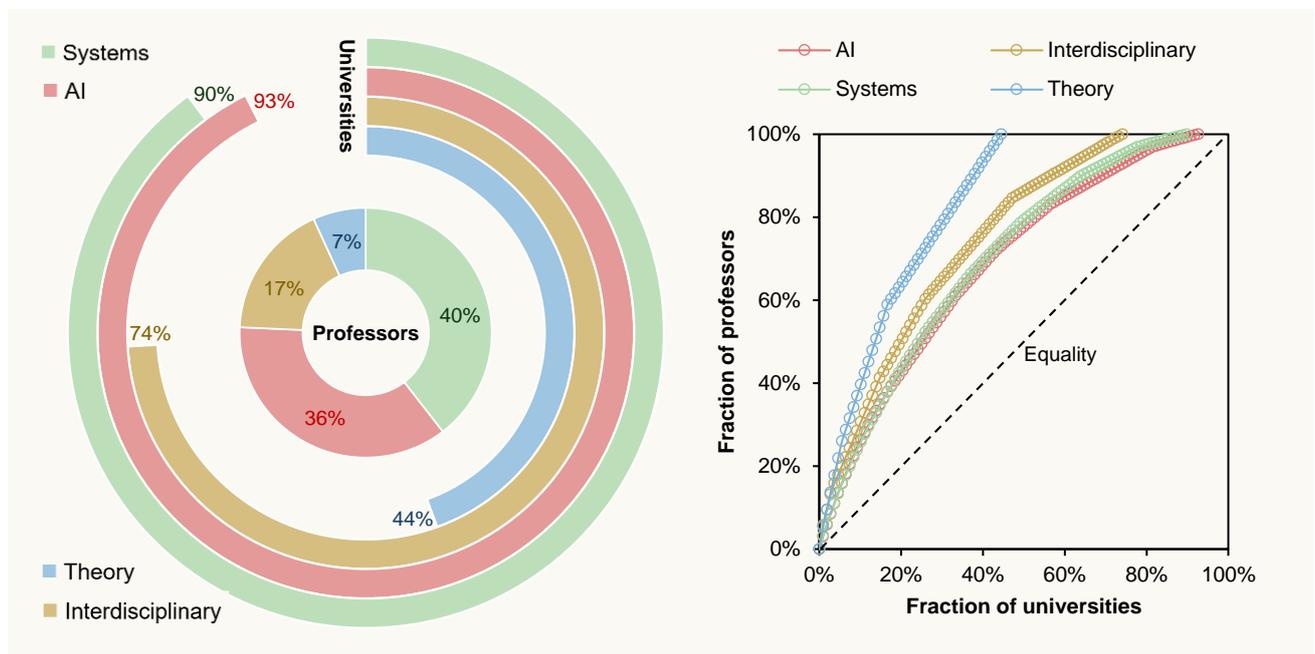

**Fig. 2** Overview of distribution by research field: numbers (left) vs. inequality (right)

Fig. 2 summarizes the specialties of 1078 US-based Chinese-descent computer science professors, along with their distribution among 108 universities by research field. On the whole, AI and Systems dominate the fields of research, involving approximately 76% of the professors and covering more than 90% of the universities. In contrast, only 24% of the professors in total are engaged in Interdisciplinary and Theory, and they are distributed among much fewer universities. Furthermore, it can be found that inequality in employment across the discipline of computer science as a whole, as measured by the Gini coefficient (denoted as $G$, with $G=0$ representing perfect equality and $G=1$ maximal inequality), is ($G=0.32$) smaller than the inequality in any of the four fields. Of these fields, employment inequality is lowest in AI ($G=0.39$), followed by Systems ($G=0.41$) and then Interdisciplinary ($G=0.51$), and highest in Theory ($G=0.66$).

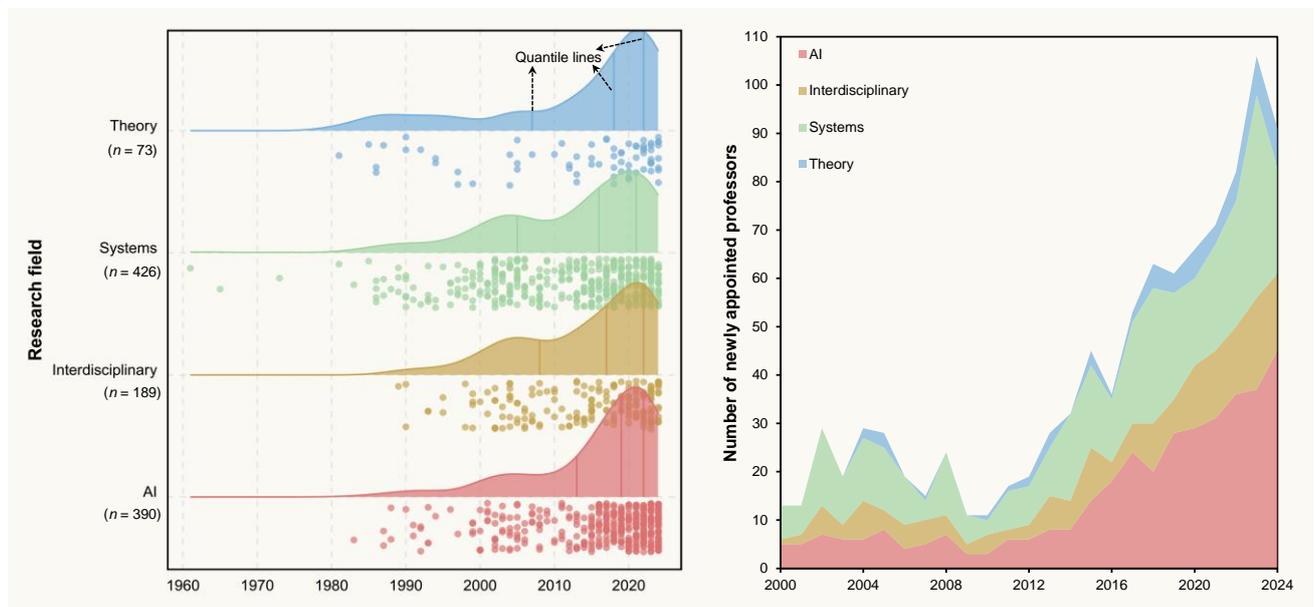

**Fig. 3** Trends of annual employment by research field: unstacked (left) vs. stacked (right)

To reach a dynamic understanding of the focal professoriate's careers, Fig. 3 shows the annual employment trends by research field. It is clear that the professors with the longest careers belong to Systems, having held their faculty positions from the 1960s onwards. Also, AI (Systems) is the youngest (oldest) field, as indicated by the quartiles, meaning that professors in this field entered US academia later (earlier) than their counterparts in any other field. In addition, it is worth noting that since 2010, the proportion of newly appointed professors working on AI has gradually increased as the year progresses. Remarkably, nearly 50% of the newly appointed professors for 2024 are engaged in AI, the same share as in the other three fields combined. In short, AI is the field that has retained a larger number of US-based Chinese-descent computer science professors in recent years, which is generally consistent with the fact that the AI spring began in the late 2010s before gaining international prominence in the early 2020s (Manyika & Bughin, 2019, October 14;

Tattershall, Nenadic, & Stevens, 2020).

Further, Fig. 4 presents the trends in average time between doctorate and employment by research field. Note that only those who started as an assistant professor are included in this illustration. Basically, over the past two decades, the time span from earning a doctorate to getting a faculty position (i.e., the preparation time for a faculty position) has lengthened in all four fields for the focal professoriate, partly due to a growing shortage in faculty positions for new doctors and a rapid explosion in postdoc positions at the same time (Gowder, 2024, February 15; Jones, 2013). In addition, the professors working on Interdisciplinary (Systems) always require a longer (shorter) preparation time to start their faculty careers following the completion of their doctoral degrees. In 2024, for example, the preparation time to attain a faculty position is 1.30 years on average for professors specializing in Systems, while that for those specializing in Interdisciplinary is 2.27 years, about 1.75 times the length of the former. Moreover, here comes a fairly counterfactual observation: in three of the four fields (except Systems), the preparation time for a faculty position decreased significantly around the year 2004, whereas previous studies suggested that it should rise at that time (Jones, 2013).

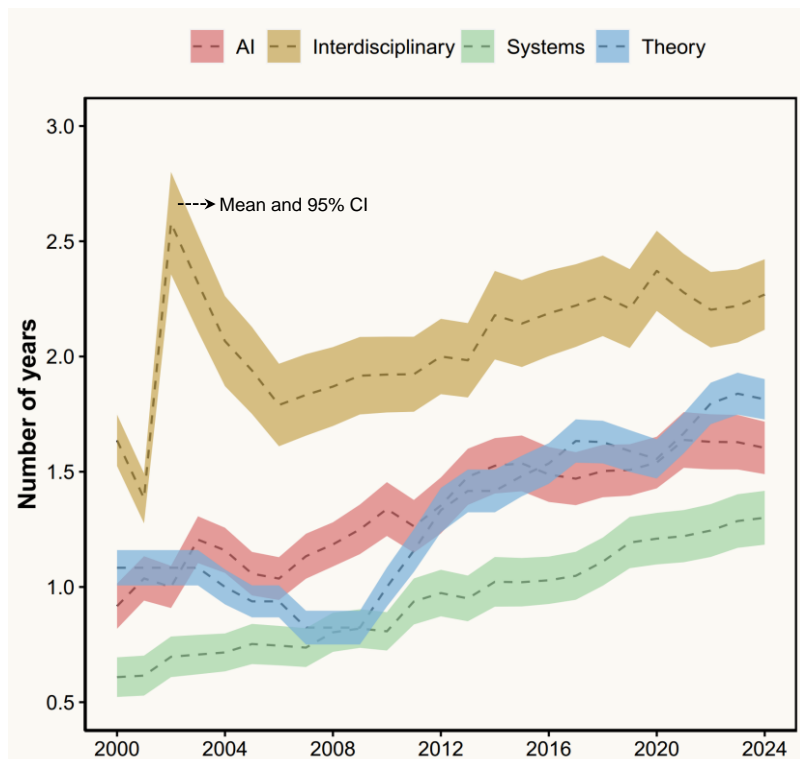

**Fig. 4** Trends of average time between doctorate and employment by research field
Note: only those who started as an assistant professor are included.

**Professors without study experience at US universities suffering more**

Table 1 summarizes the education backgrounds of 1078 US-based Chinese-descent computer science professors, in the form of cross-tabulation of doctorate and baccalaureate. Notably, the majority (72%) of the professors possess a US doctorate and a non-US baccalaureate, and there are only 0.2% on the contrary, i.e., holding a non-US doctorate and a US baccalaureate. More importantly, 90% of the professors received their doctoral degrees from US universities, and 81% completed their bachelor's degrees at universities outside the US. This pair of large percentages aligns with the realities that the US is indeed leveraging global education investments through the enrollment of foreign students (Ganguli et al., 2020; Rovito et al., 2021) and that it has long been the premier destination for these international students seeking job opportunities and building a better career path (Han, Stocking, Gebbie, & Appelbaum, 2015). Furthermore, as can be drawn from Fig. 5, inequality in the production of bachelor's degrees among non-US universities is marked ($G=0.81$) and far higher than that among US universities ($G=0.50$), and also much higher than the inequalities in the production of doctoral degrees among US universities ($G=0.59$) and among non-US universities ($G=0.42$). One last point worth mentioning is that 88% of the non-US baccalaureates are awarded by universities in mainland China (refer to Table A4 for more details, in **Appendix**).

**Table 1** Overview of distribution by education background

|  |  |  | Doctorate | | Total |
|---|---|---|---|---|---|
|  |  |  | US | Non-US |  |
| Baccalaureate | US | Count | 150 | 2 | 152 |
|  |  | % within Baccalaureate | 99% | 1% | 100% |
|  |  | % within Doctorate | 15% | 2% | 14% |
|  | Non-US | Count | 775 | 93 | 868 |
|  |  | % within Baccalaureate | 89% | 11% | 100% |
|  |  | % within Doctorate | 80% | 86% | 81% |
|  | Unknown | Count | 45 | 13 | 58 |
|  |  | % within Baccalaureate | 78% | 22% | 100% |
|  |  | % within Doctorate | 5% | 12% | 5% |
| Total |  | Count | 970 | 108 | 1078 |
|  |  | % within Baccalaureate | 90% | 10% | 100% |
|  |  | % within Doctorate | 100% | 100% | 100% |

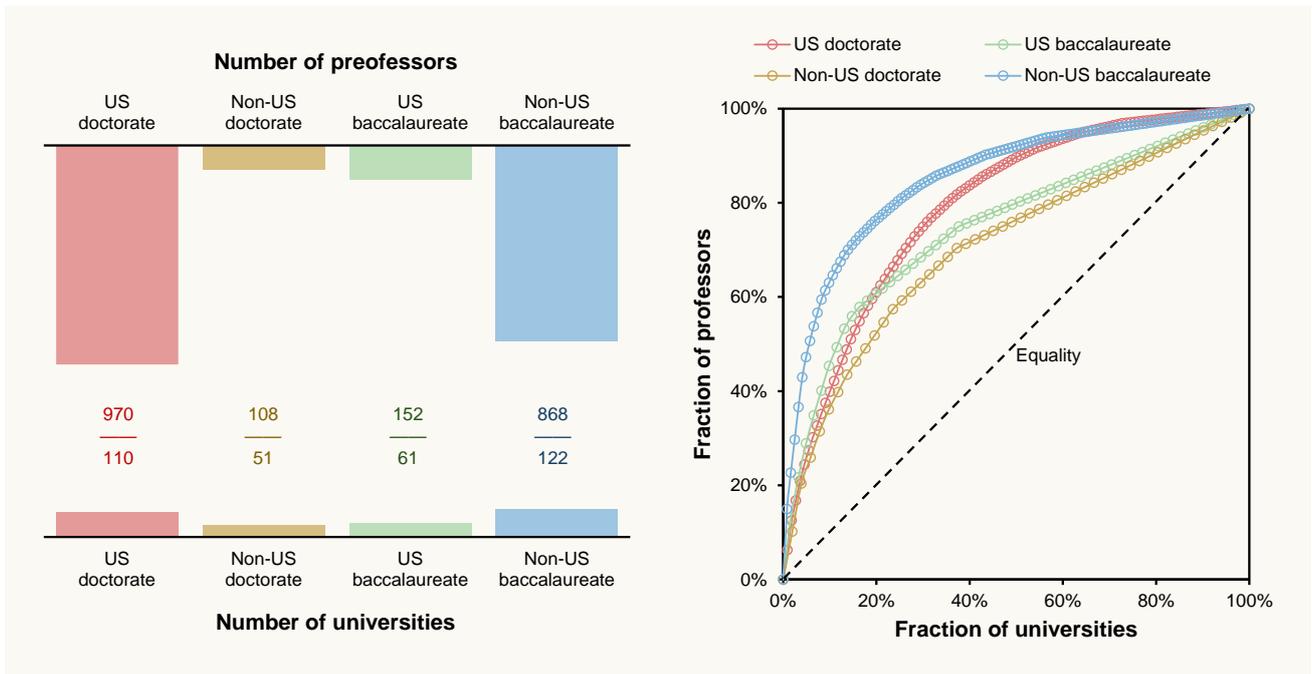

**Fig. 5** Overview of distribution by academic degree: numbers (left) vs. inequality (right)

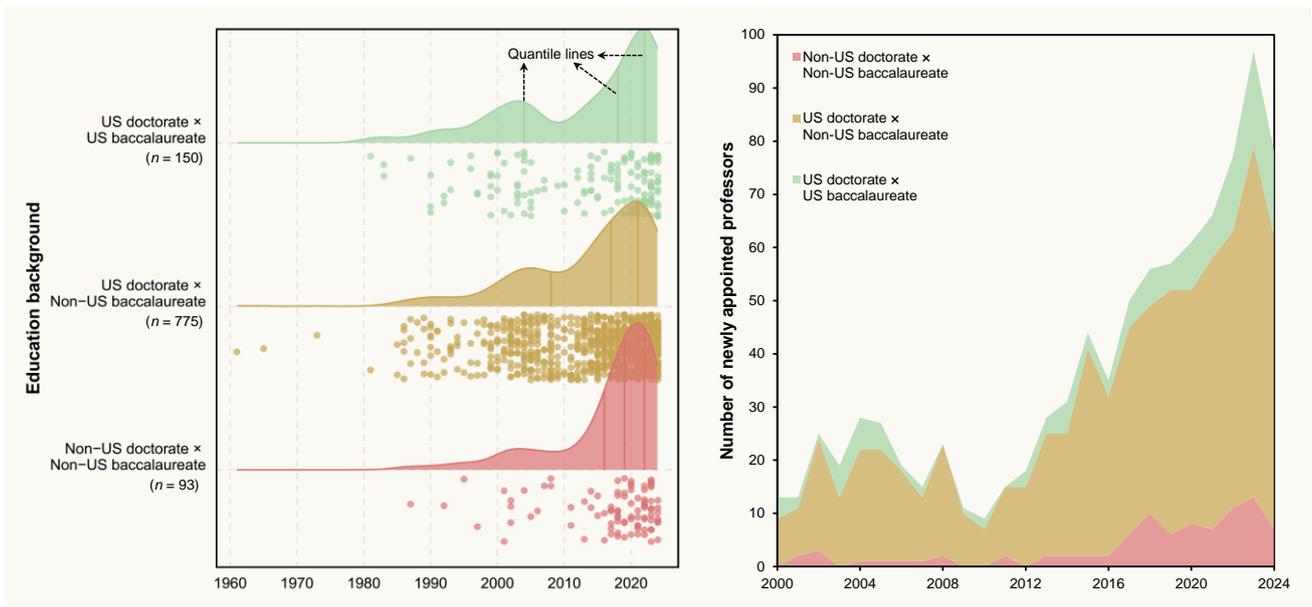

**Fig. 6** Trends of annual employment by education background: unstacked (left) vs. stacked (right)

Note: the professors holding a non-US doctorate and a US baccalaureate as well as those without the annotations of bachelor's alma maters are excluded.

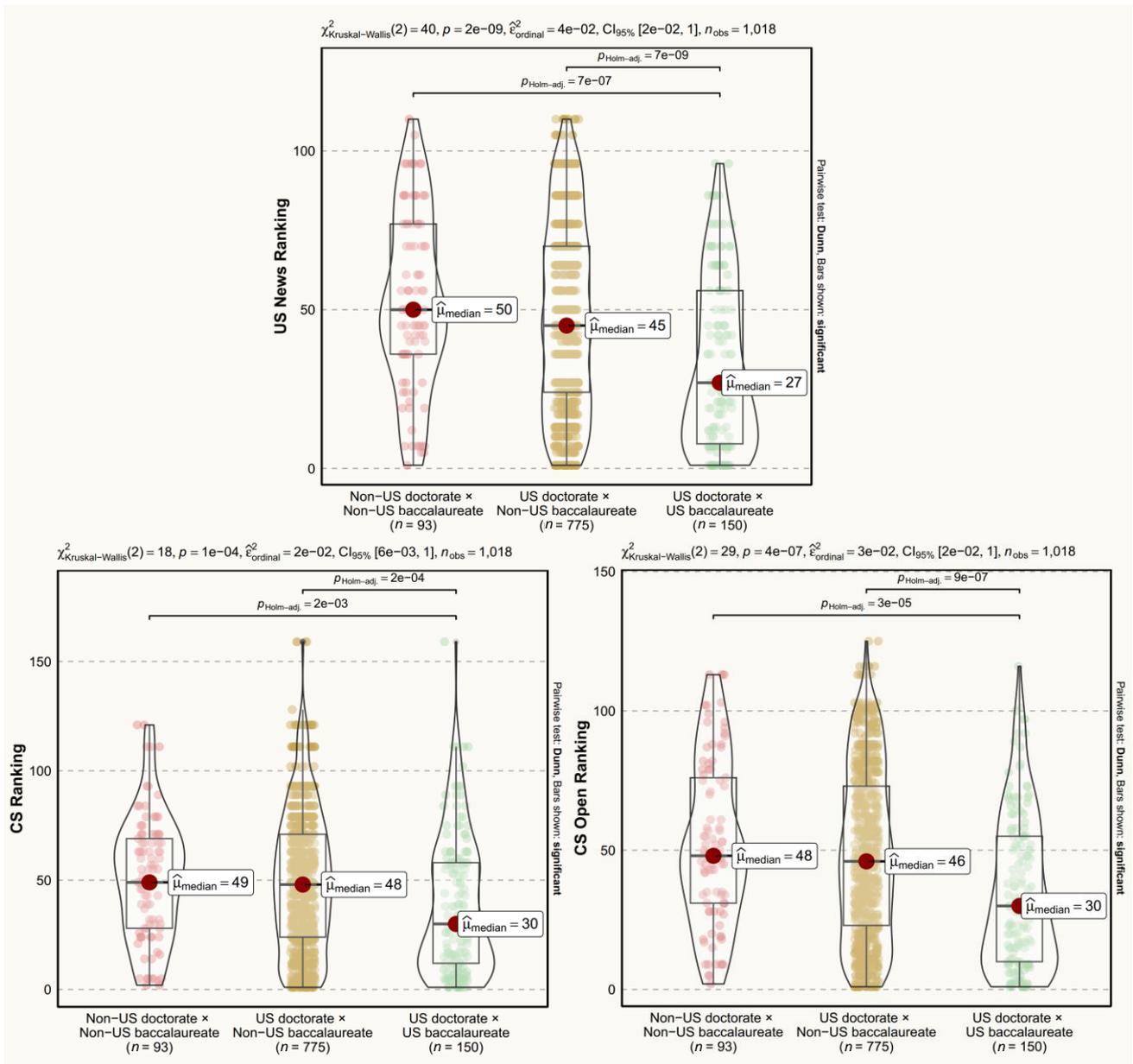

**Fig. 7** Comparison of universities that employ professors from different education backgrounds, based on three ranking systems: US News, CS, and CS Open

Note: the professors holding a non-US doctorate and a US baccalaureate as well as those without the annotations of bachelor's alma maters are excluded.

To more intuitively understand the interaction between the focal professoriate's education backgrounds and employment dynamics, Fig. 6 shows the annual employment trends by education background. Note that the professors holding a non-US doctorate and a US baccalaureate and those without the annotations of bachelors' alma maters are excluded from this illustration. It is obvious that the distribution of the newly appointed professors lacking a degree from the US higher education system is typically left-skewed, and 75% of them were appointed within the past few years. Comparatively, the distribution of those whose bachelor's and doctoral degrees are both awarded by US universities is bimodal, and quite a proportion of them entered US academia before 2000, as indicated by the first quartiles. Therefore, it is plausible to infer that the professors who had never studied at US universities suffered more from China-US tensions, given that they account for only a small percentage (≈9%) of the whole population. As a side note, Fig. 7 highlights the long-term benefits of the American undergraduate experience on career development for computer science professors of Chinese descent in the US. It can be seen that the professors holding a US doctorate and a US baccalaureate are employed by universities with, on average, higher rankings than those with non-US bachelor's or doctoral degrees. This observation is largely congruent with the findings by Xie (2023), which points to significant and lasting penalties due to non-US education for STEM workers from China to the US.

**Situational challenges emerging for women faculty in computer science**

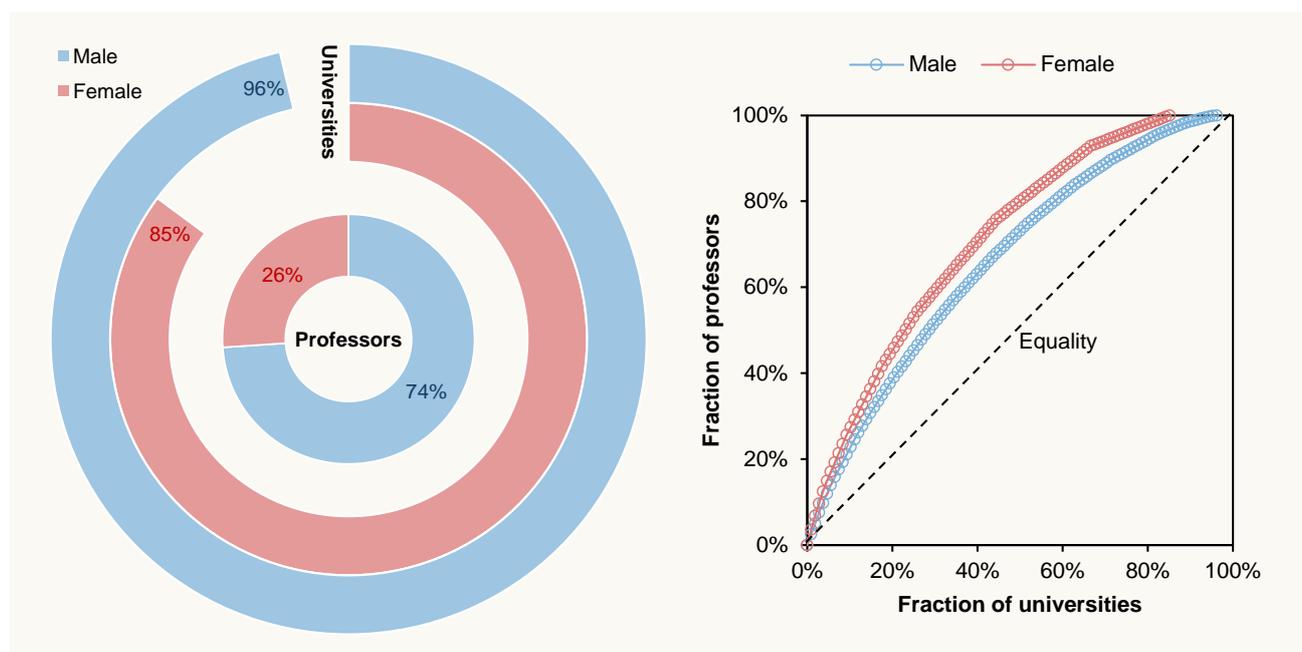

**Fig. 8** Overview of distribution by gender: numbers (left) vs. inequality (right)

Fig. 8 presents the gender statistics of 1078 US-based Chinese-descent computer science professors, along with their distribution among 108 universities by gender. Firstly, it can be found that the ratio of men to women is 2.84 vs. 1, much higher than that (1.78 vs. 1) reported by Wapman, Zhang, Clauset, and Larremore (2022) who examined tenured and tenure-track faculty employed in the years 2011-2020 at 368 PhD-granting universities in the US. Moreover, inequality in employment for the male professors ($G$=0.33) is smaller than that for their female counterparts ($G$=0.43). Generally speaking, employment inequality is indeed gendered. Surprisingly, however, as can be observed from Fig. 9, the annual employment trends for male and female professors are almost identical, indicating that the two groups are very close in average career age (see Fig. A1 for more details, in **Appendix**). That is to say, China-US tensions seemed to exert similar effects on both groups. But is this really the case?

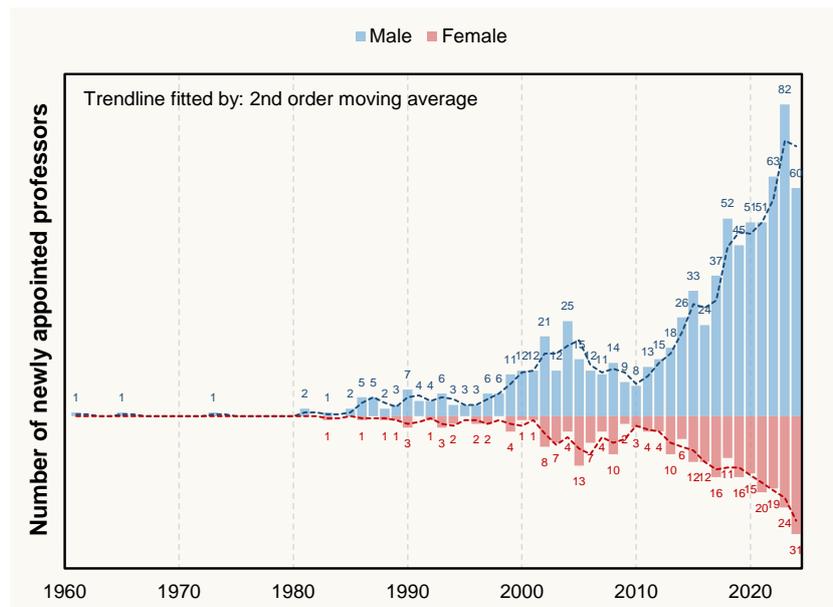

**Fig. 9** Trends of annual employment by gender

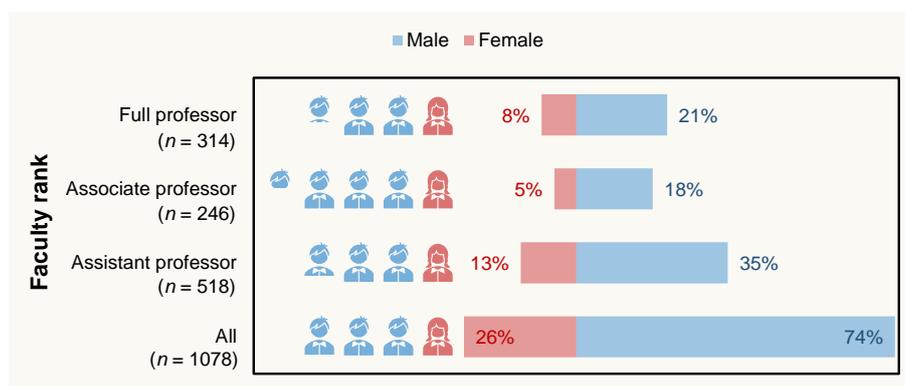

**Fig. 10** Gender ratios by faculty rank

To further investigate whether the focal professoriate's reactions to China-US tensions are gendered, Fig. 10 shows the gender ratios by faculty rank. It is evident that the ratio of men to women is most imbalanced among associate professors (3.47 vs. 1), followed by assistant professors (2.73 vs. 1) and finally full professors (2.61 vs. 1). Also, associate professors make up the smallest proportion (23%) of the whole population than do full professors (accounting for 29%). In addition, unlike the findings by Laberge et al. (2022), the female professors are employed by universities with comparable rankings to their male counterparts (see Fig. A2 for more details, in **Appendix**). Together, the high imbalance in the gender ratio among associate professors cannot be explained by faculty attrition alone, despite the fact that the underlying reasons for why men and women leave a faculty position are gendered in the US (Spoon et al., 2023). It is possible to assume that female associate professors are more vulnerable to the geopolitical tensions and more likely to be pushed out of US academia than male ones. By the way, it can be seen that 48% of the professors are in junior faculty positions, which tallies with the observation in Fig. 1, i.e., the focal professoriate has been experiencing large-scale turnover.

Finally, Fig. 11 compares the fractions of male and female full professors who are elected as ACM fellows and IEEE fellows, in order to reveal the hierarchy of the focal professoriate in more detail. It can be seen that the number of IEEE fellows is nearly twice that of ACM fellows. But when it comes to gender distribution, the ratio of men to women among IEEE fellows is 3.29 vs. 1, while that among ACM fellows is 2.22 vs. 1. The former (latter) is greater (smaller) than the gender ratio among full professors. As a matter of fact, although ACM and IEEE are both prominent professional organizations for the discipline of computer science, their focus areas are not the same——in general, ACM primarily focuses on computing (e.g., human-computer interaction, information retrieval, and programming languages), whereas IEEE concentrates on engineering (e.g., electrical engineering, electronics, and telecommunications). This finding adds to the existing knowledge on gender stratification in computer science (Frachtenberg & Kaner, 2022), leaving open questions about how scientific elites respond to geopolitical tensions.

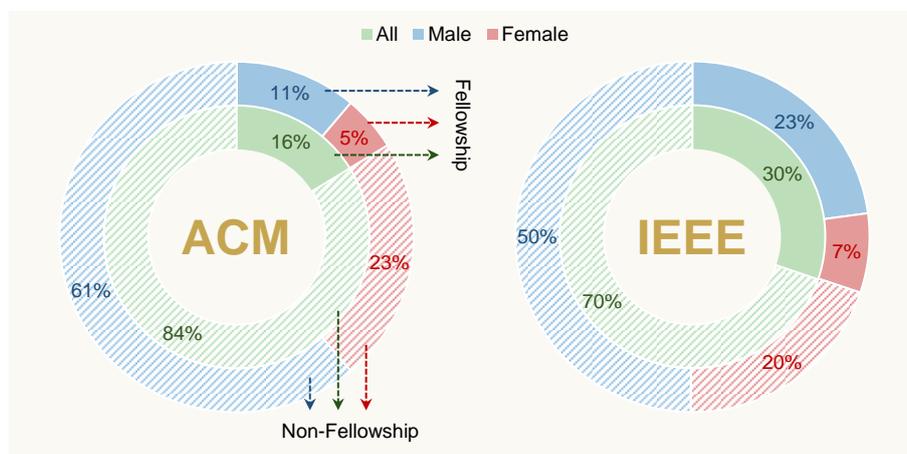

**Fig. 11** Fractions of full professors who are elected as ACM fellows and IEEE fellows

## Discussion and conclusions

The geopolitical tensions between China and the US are significantly shifting the dynamics of the American scientific workforce. To further understand this circumstance, this study selects the discipline of computer science as a representative case for empirical investigations. The findings presented in this paper suggest that China-US tensions have made it more difficult for the US higher education system to retain valuable computer science professors of Chinese descent, particularly those in their mid-to late career stages, and that almost half of the existing professors have less than seven years of faculty experience. Additionally, the deterioration in faculty retention varies across fields of research, education backgrounds, and gender groups. To be specific, among the professors we are concerned about, those who do not work on AI or Systems, those who lack study experience at US universities, and those who are women, are underrepresented, albeit in different forms and to varying degrees. In summary, the focal professoriate has not only shrunk in size, as has been widely reported, but also lost some of its diversity in structure.

The structural change in the focal professoriate may trigger butterfly effects, since evidence indicates that scientists of Chinese descent play as a vital link for scientific cooperation and education exchange globally, especially between China and the US (Cao, Baas, Wagner, & Jonkers, 2020; Flynn et al., 2024). As things stand at present, many Chinese-descent computer science professors migrating out of the US are senior faculty who can already have established a strong research base and maintained a reliable collaboration network in US academia. Their departure could deal a huge blow to their collaborators, the students they mentored, and the organizations they served in the US, if the geopolitical tensions persisted or escalated and ultimately led to a severance of all of the above ties. Moreover, they could even be completely knocked out, i.e., they could exit the scientific community forever, if the institutions (located outside the US) accepting them were unable to help them rebuild their careers in time. No matter which possibility becomes a reality, it will be a great loss for the realm of science, definitely. One last point: special actions should be taken immediately to diversify the focal professoriate (e.g., along gender, schooling, and expertise lines), because it has been proven that fostering diversity and inclusion in an academic environment contributes to shaping what discoveries are made and who makes them.

The findings presented in this paper should be interpreted with caution due to the following reasons. First of all, the sample created in this study is cross-sectional, so that causal interpretations of any findings should have been rejected in general. The few inferences that link any findings to China-US tensions are either supported by previous studies or made purely out of rational intuition, such as the inference drawn from the seemingly illogical proportion of female associate professors. Admittedly, these inferences are debatable and remain to be verified in the future. Second, only tenured and tenure-track faculty at first-class universities were sampled in this study owing to limited manpower. So it is unclear whether the findings can be extended to, for example, non-tenure track faculty or faculty at all tiers of universities. Last but not least, please do not take the exhibited data as gospel. The sample was created based on information collected from whatever sources I could find on the web; I'm positive that there are mistakes.

# Appendix

**Table A1** Universities in the US: origins of the sample in this study

| No | University | Ranking System | | |
|---|---|---|---|---|
| | | US News | CS | CS Open |
| 1 | Carnegie Mellon University | 1 | 1 | 1 |
| 2 | Massachusetts Institute of Technology | 1 | 4 | 2 |
| 3 | University of California, Berkeley | 1 | 7 | 3 |
| 4 | Stanford University | 1 | 10 | 3 |
| 5 | University of Illinois Urbana-Champaign | 5 | 2 | 5 |
| 6 | Cornell University | 6 | 9 | 7 |
| 7 | University of Washington | 7 | 8 | 6 |
| 8 | Georgia Institute of Technology | 7 | 5 | 9 |
| 9 | University of Texas at Austin | 7 | 15 | 10 |
| 10 | University of Michigan | 10 | 6 | 8 |
| 11 | Princeton University | 10 | 19 | 11 |
| 12 | Columbia University | 12 | 17 | 15 |
| 13 | University of California, San Diego | 13 | 3 | 11 |
| 14 | University of Wisconsin-Madison | 13 | 17 | 14 |
| 15 | University of California, Los Angeles | 13 | 23 | 16 |
| 16 | California Institute of Technology | 13 | 84 | 38 |
| 17 | University of Maryland | 17 | 11 | 13 |
| 18 | Harvard University | 17 | 39 | 20 |
| 19 | University of Pennsylvania | 19 | 16 | 17 |
| 20 | Purdue University | 19 | 14 | 18 |
| 21 | University of Southern California | 21 | 22 | 22 |
| 22 | Yale University | 21 | 38 | 30 |
| 23 | Duke University | 21 | 30 | 31 |
| 24 | University of Massachusetts Amherst | 24 | 24 | 19 |
| 25 | University of Chicago | 24 | 26 | 23 |
| 26 | Johns Hopkins University | 24 | 53 | 36 |
| 27 | New York University | 27 | 21 | 20 |
| 28 | Brown University | 27 | 46 | 24 |
| 29 | Northwestern University | 27 | 33 | 25 |
| 30 | University of California, Irvine | 27 | 27 | 26 |
| 31 | Ohio State University | 27 | 37 | 27 |
| 32 | Rice University | 27 | 52 | 28 |
| 33 | Northeastern University | 27 | 12 | 33 |
| 34 | University of California, Santa Barbara | 27 | 29 | 39 |
| 35 | University of North Carolina at Chapel Hill | 27 | 159 | 64 |
| 36 | University of Virginia | 36 | 43 | 31 |
| 37 | University of Minnesota | 36 | 49 | 35 |

| No | University | Ranking System | | |
|---|---|---|---|---|
| | | US News | CS | CS Open |
| 38 | Virginia Tech | 36 | 55 | 37 |
| 39 | University of California, Davis | 36 | 49 | 42 |
| 40 | Pennsylvania State University | 40 | 32 | 34 |
| 41 | University of Colorado Boulder | 40 | 53 | 40 |
| 42 | Rutgers University | 42 | 28 | 28 |
| 43 | University of Utah | 42 | 34 | 44 |
| 44 | Washington University in St. Louis | 42 | 60 | 48 |
| 45 | Stony Brook University | 45 | 24 | 43 |
| 46 | Arizona State University | 45 | 49 | 46 |
| 47 | Texas A&M University | 45 | 39 | 47 |
| 48 | University of Florida | 45 | 71 | 55 |
| 49 | Boston University | 45 | 44 | 58 |
| 50 | North Carolina State University | 50 | 44 | 40 |
| 51 | Indiana University Bloomington | 50 | 55 | 45 |
| 52 | University of Rochester | 50 | 60 | 49 |
| 53 | Dartmouth College | 50 | 71 | 53 |
| 54 | University of California, Santa Cruz | 50 | 42 | 62 |
| 55 | Vanderbilt University | 50 | 75 | 73 |
| 56 | University of Notre Dame | 56 | 63 | 52 |
| 57 | University of Pittsburgh | 56 | 63 | 54 |
| 58 | Michigan State University | 56 | 63 | 55 |
| 59 | University of California, Riverside | 56 | 30 | 58 |
| 60 | Tufts University | 56 | 89 | 68 |
| 61 | University at Buffalo | 61 | 48 | 57 |
| 62 | University of Illinois Chicago | 61 | 46 | 61 |
| 63 | University of Arizona | 61 | 84 | 65 |
| 64 | Iowa State University | 64 | 75 | 51 |
| 65 | George Mason University | 64 | 34 | 66 |
| 66 | University of Texas at Dallas | 64 | 58 | 69 |
| 67 | University of Oregon | 64 | 93 | 72 |
| 68 | Rensselaer Polytechnic Institute | 64 | 93 | 74 |
| 69 | Emory University | 64 | 89 | 97 |
| 70 | Oregon State University | 70 | 57 | 50 |
| 71 | College of William & Mary | 70 | 79 | 60 |
| 72 | University of Central Florida | 70 | 58 | 63 |
| 73 | University of Delaware | 70 | 71 | 75 |
| 74 | Georgetown University | 70 | 79 | 86 |
| 75 | George Washington University | 70 | 102 | 89 |
| 76 | Rochester Institute of Technology | 70 | 63 | 96 |
| 77 | University of Maryland, Baltimore County | 77 | 111 | 67 |
| 78 | Case Western Reserve University | 77 | 111 | 71 |

| No | University | Ranking System | | |
|---|---|---|---|---|
| | | US News | CS | CS Open |
| 79 | Toyota Technological Institute at Chicago | 77 | 84 | 76 |
| 80 | Florida State University | 77 | 84 | 78 |
| 81 | Stevens Institute of Technology | 77 | 79 | 79 |
| 82 | Syracuse University | 77 | 111 | 80 |
| 83 | University of Connecticut | 77 | 69 | 81 |
| 84 | University of Iowa | 77 | 111 | 91 |
| 85 | Colorado School of Mines | 77 | 93 | 116 |
| 86 | University of Nebraska-Lincoln | 86 | 111 | 70 |
| 87 | Washington State University | 86 | 89 | 77 |
| 88 | Worcester Polytechnic Institute | 86 | 75 | 82 |
| 89 | Clemson University | 86 | 79 | 82 |
| 90 | Drexel University | 86 | 102 | 85 |
| 91 | University of Tennessee | 86 | 93 | 92 |
| 92 | Lehigh University | 86 | 93 | 92 |
| 93 | New Jersey Institute of Technology | 86 | 79 | 100 |
| 94 | University of Georgia | 86 | 84 | 102 |
| 95 | Auburn University | 86 | 121 | 103 |
| 96 | University of North Carolina at Charlotte | 96 | 93 | 84 |
| 97 | Binghamton University | 96 | 71 | 87 |
| 98 | University of Texas at Arlington | 96 | 67 | 88 |
| 99 | Illinois Institute of Technology | 96 | 102 | 95 |
| 100 | Temple University | 96 | 93 | 101 |
| 101 | University of California, Merced | 96 | 67 | 113 |
| 102 | Colorado State University | 96 | 142 | 120 |
| 103 | Naval Postgraduate School | 96 | 159 | 125 |
| 104 | Rutgers University-Newark | 96 | N/A | 161 |
| 105 | University of New Mexico | 105 | 128 | 90 |
| 106 | University of Kansas | 105 | 121 | 99 |
| 107 | Wayne State University | 110 | 93 | 94 |
| 108 | Brandeis University | 110 | 121 | 98 |

Source: https://drafty.cs.brown.edu/csopenrankings/ (accessed July 1, 2024).

**Table A2** Taxonomy of computer science fields and their subfields

| Field | Subfield |
|---|---|
| AI | Artificial intelligence |
| | Computer vision |
| | Machine learning |
| | Natural language processing |
| | The Web & information retrieval |
| Interdisciplinary | Comp. bio & bioinformatics |
| | Computer graphics |
| | Computer science education |
| | Economics & computation |
| | Human-computer interaction |
| | Robotics |
| | Visualization |
| Systems | Computer architecture |
| | Computer networks |
| | Computer security |
| | Databases |
| | Design automation |
| | Embedded & real-time systems |
| | High-performance computing |
| | Mobile computing |
| | Measurement & perf. analysis |
| | Operating systems |
| | Programming languages |
| | Software engineering |
| Theory | Algorithms & complexity |
| | Cryptography |
| | Logic & verification |

Source: https://csrankings.org/ (accessed July 1, 2024).

**Table A3** Pearson's correlation matrix of annotated data

|                          | A       | B       | C       | D       | E      | F      |
|--------------------------|---------|---------|---------|---------|--------|--------|
| A. gender                | 1.0000  |         |         |         |        |        |
| B. faculty rank          | 0.0029  | 1.0000  |         |         |        |        |
| C. research field        | −0.0453 | 0.0924  | 1.0000  |         |        |        |
| D. education background  | −0.0274 | −0.0369 | −0.0456 | 1.0000  |        |        |
| E. academic age          | −0.0295 | 0.8307  | 0.1227  | −0.0164 | 1.0000 |        |
| F. career age            | −0.0252 | 0.8236  | 0.1546  | −0.0868 | 0.9531 | 1.0000 |

Note: the professors without the annotations of bachelor's alma maters were not included in analyses of education background but were included in all other analyses.

Coding:

A. gender: male=0, female=1;

B. faculty rank: assistant professor=0, associate professor=1, full professor=2;

C. research field: AI=0, Interdisciplinary=1, Systems=2, Theory=3;

D. education background: US doctorate×US baccalaureate=0,
US doctorate×Non-US baccalaureate=1,
Non-US doctorate×US baccalaureate=2,
Non-US doctorate×Non-US baccalaureate=3;

E. academic age=2024−year of receiving the doctorate;

F. career age=2024−year of starting on the faculty career in US academia.

**Table A4** Universities in mainland China: some origins of non-US baccalaureates

| No | University |
|---|---|
| 1 | Tsinghua University |
| 2 | Shanghai Jiao Tong University |
| 3 | Peking University |
| 4 | University of Science and Technology of China |
| 5 | Zhejiang University |
| 6 | Huazhong University of Science and Technology |
| 7 | Fudan University |
| 8 | Nanjing University |
| 9 | Beihang University |
| 10 | Wuhan University |
| 11 | Beijing Institute of Technology |
| 12 | Xidian University |
| 13 | Harbin Institute of Technology |
| 14 | Northwestern Polytechnical University |
| 15 | Xi'an Jiaotong University |
| 16 | Shandong University |
| 17 | Beijing University of Posts and Telecommunications |
| 18 | Sichuan University |
| 19 | Tongji University |
| 20 | Nanjing University of Posts and Telecommunications |
| 21 | University of Electronic Science and Technology of China |
| 22 | Tianjin University |
| 23 | Jilin University |
| 24 | Nankai University |
| 25 | Dalian University of Technology |
| 26 | Beijing Jiaotong University |
| 27 | South China University of Technology |
| 28 | Beijing Normal University |
| 29 | Nanjing University of Aeronautics and Astronautics |
| 30 | Sun Yat-sen University |
| … | … |

Note: sort by the number of professors in descending order.

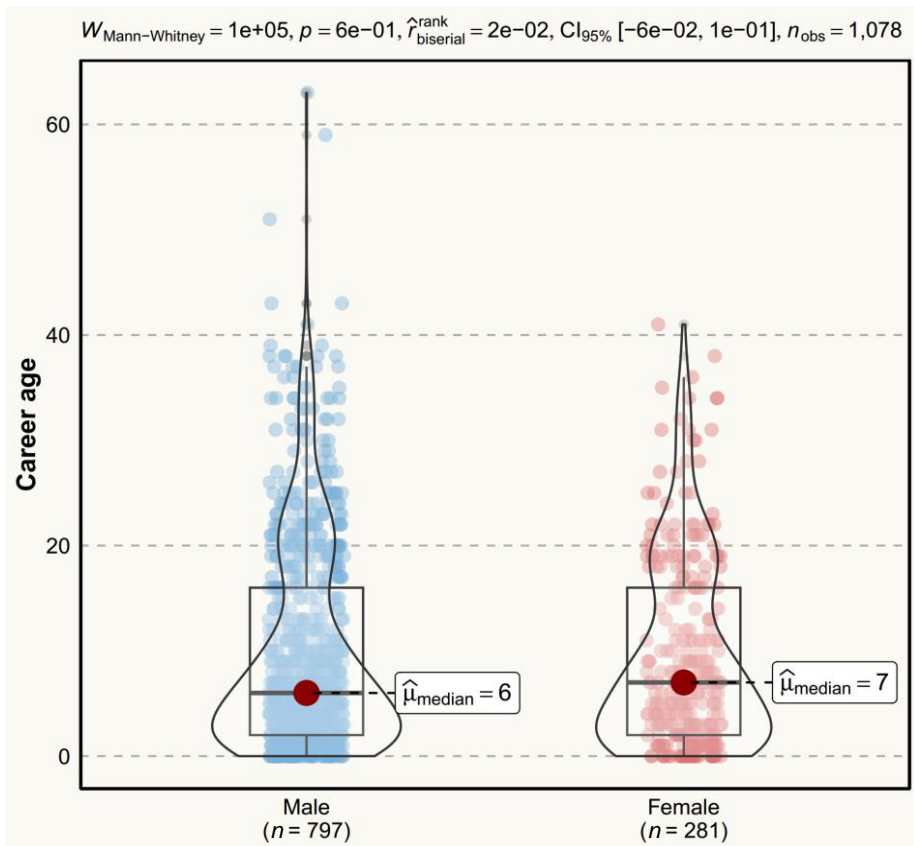

**Fig. A1** Comparison of career ages: male vs. female

Note: career age=2024−year of starting on the faculty career in US academia.

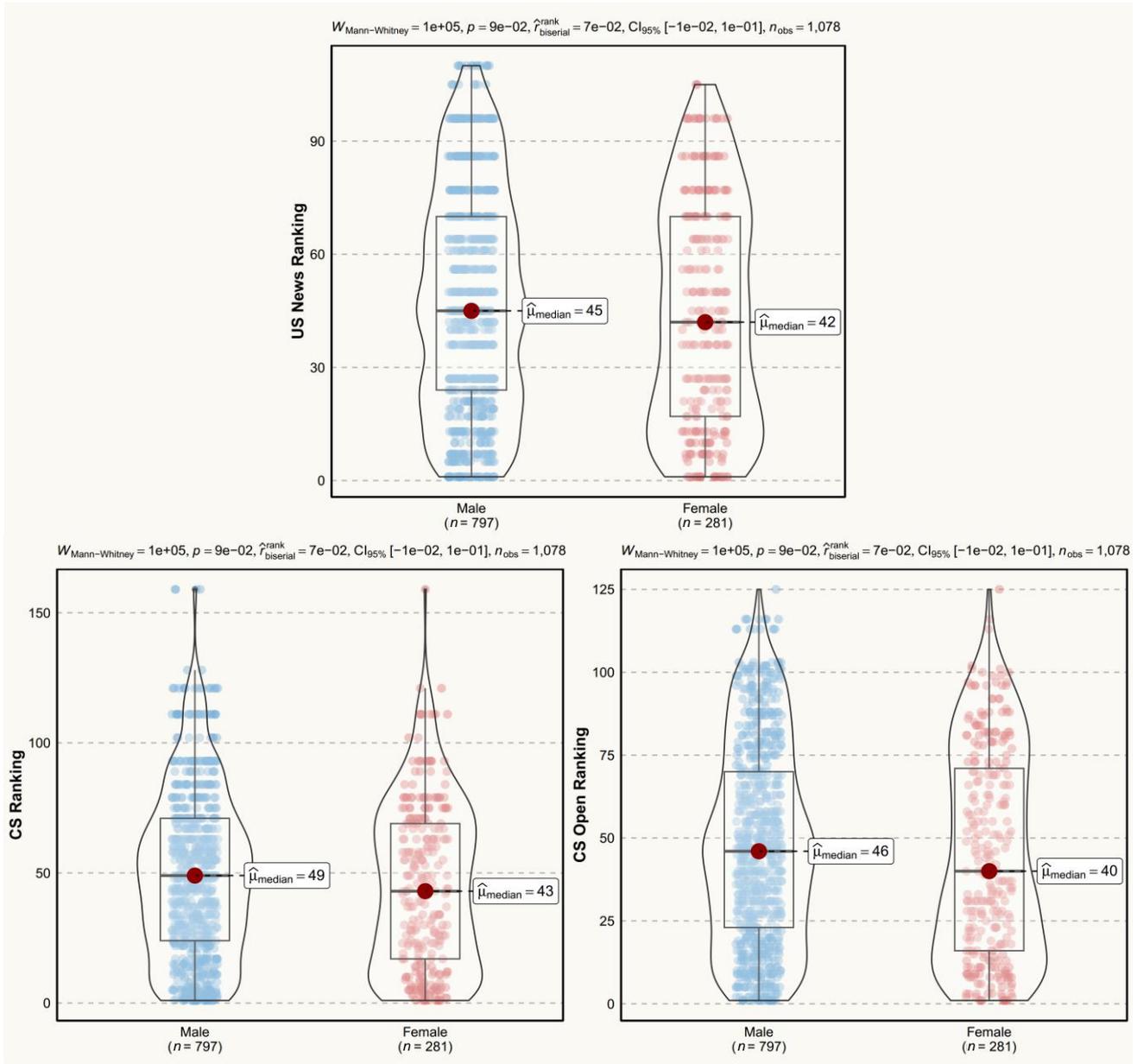

**Fig. A2** Comparison of universities that employ male and female professors, based on three ranking systems: US News, CS, and CS Open